\documentclass[twocolumn,showpacs,aps,prl,letterpaper,superscriptaddress]{revtex4}

\usepackage{graphicx}
\usepackage{dcolumn}
\usepackage{amsmath}
\usepackage{epsfig}
\usepackage{subfigure}

\input babarsym

\newcommand{\BABARPubYear}    {04}
\newcommand{\BABARPubNumber}  {021}

\newcommand{\SLACPubNumber} {10568}

\newcommand{\Btaunu}  {\ensuremath{\Bm \to \taum \bar{\nut}\xspace}}

\newcommand{\BRBtaunu}{\ensuremath{\BR(\Btaunu)}\xspace}

\newcommand{\btodlnux}{\ensuremath{\Bp \to \Dzb \ell^{+} \nu_{\ell} X}}

\newcommand{\thelimit}{\ensuremath{4.2 \times 10^{-4}}}

\def\figurebox#1#2#3{%
    \def\arg{#3}%
    \ifx\arg\empty
    {\hfill\vbox{\hsize#2\hrule\hbox to #2{\vrule\hfill\vbox to #1{\hsize#2\vfill}\vrule}\hrule}\hfill}%
    \else
    {\hfill\epsfbox{#3}\hfill}%
    \fi}

\begin{document}


\preprint{\babar-PUB-\BABARPubYear/\BABARPubNumber} 
\preprint{SLAC-PUB-\SLACPubNumber} 

\begin{flushleft}
\babar-PUB-\BABARPubYear/\BABARPubNumber\\
SLAC-PUB-\SLACPubNumber\\
\end{flushleft}

\title{
{\large \bf  
A Search for the Rare Leptonic Decay \boldmath{\Btaunu}} 
}


%
\author{B.~Aubert}
\author{R.~Barate}
\author{D.~Boutigny}
\author{F.~Couderc}
\author{J.-M.~Gaillard}
\author{A.~Hicheur}
\author{Y.~Karyotakis}
\author{J.~P.~Lees}
\author{V.~Tisserand}
\author{A.~Zghiche}
\affiliation{Laboratoire de Physique des Particules, F-74941 Annecy-le-Vieux, France }
% ...
\collaboration{The \babar\ Collaboration}
\noaffiliation

\date{\today} 

\begin{abstract}
We present a search for the decay \Btaunu\ in a sample of $88.9 \times
10^{6}$ $\BB$ pairs recorded with the \babar\ detector at the SLAC
\B-Factory. 
One of the two \B\ mesons from the \FourS\ is
reconstructed in a hadronic or a semileptonic final state and the 
decay products of the other \B\ in the event are analyzed for
consistency with a \Btaunu\ decay.
We find no evidence of a signal and set an upper limit on the branching fraction of
$\BRBtaunu < \thelimit$ at the 90\% confidence level.
\end{abstract}


\maketitle

In the Standard Model (SM) the leptonic decay \Btaunu~\cite{bib:cc} 
proceeds via the annihilation
of the \b\ and \ubar\ quarks into a virtual \W\ boson. Its
amplitude is thus proportional to the product of the Cabibbo-Kobayashi-Maskawa 
(CKM) matrix~\cite{bib:CKM} element \Vub\ and the \B\ meson decay 
constant \fsubb.
The SM branching fraction is given by:
\begin{eqnarray}
  \BR(\Btaunu) & = & \frac{G^2_F m_B}{8 \pi} m_{\tau}^2 
        \left( 1 - \frac{m^2_{\tau}}{m^2_{B}} \right)^2 f^2_{B} \Vub
        ^2 \tau_{B}\nonumber \\
        & = & (9.3 \pm 3.9) \times 10^{-5}\:,
  \label{eq:brsm}
\end{eqnarray}
where $G_F$ is the Fermi coupling constant, $m_{\tau}$ and $m_{B}$ 
are the $\tau$ lepton and \Bub\ meson masses, and $\tau_{B}$ is the \Bm\ mean
lifetime. We have used $\tau_{B} = (1.671 \pm 0.018) \ps$,
$\Vub = (3.67 \pm 0.47)\times 10^{-4}$, and 
$\fsubb = (0.196 \pm 0.032) \gev$
(obtained from lattice QCD calculations)~\cite{bib:pdg2002}. 
The branching fractions for $\en\nueb$ and $\mun\numb$ are helicity
suppressed by factors of $\sim 10^{-8}$ and $\sim 10^{-3}$, respectively.
Physics beyond the SM, such as supersymmetry or two-Higgs doublet models, 
could enhance $\BR(\Btaunu)$ by up to a factor of five through the
introduction of a charged Higgs boson~\cite{bib:higgs}.

A search for this decay is experimentally challenging due to the
presence of at least two undetectable neutrinos in the final state.
No observation has been reported yet and the most stringent published limit on
the decay is $\BRBtaunu < 5.7 \times 10^{-4}$ at the 90\% confidence level~\cite{bib:L3}.

The data used in this analysis were recorded with the \babar\ detector
at the \pep2\ asymmetric \epem\ storage ring.
The sample consists of $88.9 \pm 1.0$ million \BB\ pairs ($81.9\invfb$)
collected at the \FourS\ resonance (``on-resonance") and $9.6\invfb$ collected
about 40\mev\ below the \BB\ threshold (``off-resonance").

The \babar\ detector is described in detail elsewhere~\cite{bib:babar}.
Detection of charged particles and measurement of their momenta are 
performed by a five-layer double-sided silicon vertex tracker
and a 40-layer drift chamber, which operate in a 1.5-T 
solenoidal magnetic field.
A detector of internally reflected Cherenkov light is
used to identify charged kaons and pions.
Photons and electrons are detected in an electromagnetic calorimeter
consisting of an array of CsI(Tl) crystals. Muons and
neutral hadrons are identified in the flux return, which is
instrumented with multiple layers of resistive plate chambers.
A {\geant}4-based~\cite{bib:geant} simulation of the \babar\ detector,
including machine backgrounds,
is used to study signal event selection and background rejection.

We first select a sample of events with one \B-meson (the {\em tag} \B)
reconstructed in a hadronic or a semileptonic  
final state. The reconstruction constrains the kinematics
and reduces the combinatorics in each event.
This is critical since
at least two neutrinos result from the \Btaunu\ decay.
All the neutral and charged particles not used for the tag 
\B\ are assumed to come from the \B-meson recoiling against it.  
We use two methods to 
search this recoil system for evidence of a \Btaunu\ signal.



\def\lp{\ensuremath{1.0\gev/c}}
\def\Dl{\ensuremath{D\ell}}
\def\cosbydl{\ensuremath{\cos{\theta_{B,\Dl}}}}
\def\eextra{\ensuremath{E_{\mathrm{extra}}}}
\def\mus90{\ensuremath{\mu_{s}^{90}}}
\def\musfit{\ensuremath{\mu_{s}^{fitted}}}
\def\slbkgfitted{\ensuremath{X \pm Y}}
\def\slbkgextrap{\ensuremath{414 \pm 23}}
\def\slsigfitted{\ensuremath{X \pm Y}}
\def\PDF{{p.d.f.}}
In our first method, we reconstruct the tag \B\ semileptonically. 
The semileptonic \B-meson, $\B_{{\rm sl}}$, is reconstructed 
as \btodlnux, 
where $\ell=e,\mu$ and $X$ can be a $\gamma$, \piz, or nothing.
We select semileptonic \B-decay events with several missing particles 
(such as neutrinos) by requiring at least one lepton with
center-of-mass (CM)
momentum ($|{\vec{p}}^{*}_{\ell}|$) above $\lp$, zero event charge, 
a ratio of the Fox-Wolfram
moments~\cite{bib:foxwolfram} $H_{2}/H_{0} < 0.9$,
and missing mass greater than $1.0\gev/c^2$.
Here, the missing mass is determined by subtracting the total energy and momentum
of all reconstructed tracks and neutrals from
the four-momentum of the \FourS\ system.
We reconstruct \Dzb\ mesons in
the modes $\Dzb \to K^+\pim,\,K^+\pim\pim\pip,\,K^+\pim\piz$, and 
$\KS\pip\pim$ and require their reconstructed masses to be
within three standard deviations of the observed mean. 
The \Dzb\ mesons are then paired with leptons with 
$|\vec{p}^{*}_{\ell}| > \lp$ to form \Dl\ candidates. 
If the \Dzb\ decay contains a charged kaon, the lepton
must have the same charge as the kaon. The \Dzb\ and lepton are
required to originate from a common vertex, but we do not
mass-constrain the vertex fit. We assume that the only
missing particle is a neutrino and calculate the cosine of the angle
between the momentum vectors of the \Dl\ candidate 
and the \B-meson,
\begin{equation}\label{eqn:cosbydl}
\cosbydl \equiv \frac{2 E^{*}_{\rm beam} E^{*}_{\Dl} - m^2_{\B}-m^2_{\Dl}}%
{2 \sqrt{E^{*2}_{\rm beam} - m^{2}_{B}} |\vec{p}^{*}_{\Dl}|}.
\end{equation}
The CM energy and momentum of the \Dl\
candidate are $E^{*}_{\Dl}$ and $\vec{p}^{*}_{\Dl}$, respectively.
The \B-meson energy is taken to be the
beam energy, $E^{*}_{\rm beam}$.
Calculated values of \cosbydl\ may lie outside the physical range
for events where the \Dl\ candidate did not arise as presumed, 
or due to detector energy and momentum resolution.
We place an asymmetric restriction on this variable, 
$-2.5 < \cosbydl < 1.1$, to admit $\Dstarzb$ states where
additional decay products are present.
If there is more than one acceptable \Dl\ candidate,
we choose the one whose \Dzb\ mass is
closest to the mean of the fitted distribution.

After identifying the $\B_{{\rm sl}}$,
the remaining particles are required to be consistent with 
$\Btaunu$ where 
$\taum \to \en \bar{\nu}_{e} \nut$ or $\mun \bar{\nu}_{\mu}
\nut$. Exactly one track with a small 
impact parameter relative to the primary vertex must remain.
The track must have $p^{*} < 1.2\gev/c$, 
and must be identified as either an electron or muon. 
We reject $\ep\en \to \taup\taum$
events by restricting the angle of the
track with respect to the event  thrust axis 
($|\cos{\theta_{\vec{p},\vec{T}}}| < 0.9$) and the minimum invariant
mass constructable from any triplet of tracks in the event
($M^{\mathrm{min}}_3>1.5\gev/c^{2}$). 
In general, continuum events tend to peak sharply at
$|\cos{\theta_{\vec{p},\vec{T}}}| = 1$ and  $\taup\taum$
events in particular tend to peak at 
values of $M^{\mathrm{min}}_3$ below the $\tau$ mass.

\begin{figure}[t]
\includegraphics[width=\linewidth]{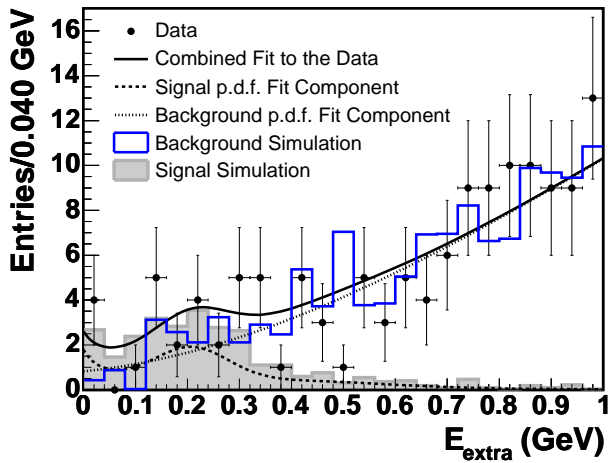}
\caption{\label{fig:sl_eextra}%
The distribution of \eextra\ 
after applying all selection criteria.
The fit to the data and its 
components are also shown.
The background is normalized to the data luminosity and
the signal simulation is normalized arbitrarily.}
\end{figure}

The signal yield in the data is determined using 
the distribution of the total energy deposited in calorimeter 
clusters (with a minimum energy of $0.020\gev$) 
by neutral particles not associated with the \Dzb\ decay 
in the semileptonic $B_{\rm sl}$ candidate, \eextra~(Fig. ~\ref{fig:sl_eextra}).
This variable peaks near zero for signal while
for background it rises with increasing \eextra. For 
$\eextra<1.0\gev$, we find from Monte Carlo
simulations a signal efficiency of 
$(4.77 \pm 0.35) \times 10^{-4}$ and a background estimate of
$124 \pm 7$ events.

The signal efficiency quoted above is determined
using a detailed signal simulation. We study the 
differences between simulation and data in the
semileptonic \B\ reconstruction, neutral-energy reconstruction,
and lepton identification to derive an efficiency correction. 
The most significant effect comes from the 
$\B_{{\rm sl}}$ reconstruction efficiency, and
is determined using a sample of  events in data and 
Monte Carlo simulations where both \B\ mesons
are reconstructed as $\B \to D \ell \nu X$. The total efficiency correction
from all sources is determined to be $0.878 \pm 0.076$, and the
corrected signal efficiency is
$(4.19 \pm 0.31_{{\rm stat}} \pm 0.36_{{\rm syst}})\times 10^{-4}$.

Probability density functions (\PDF's) are 
constructed from the \eextra\ 
distributions in signal ($F(\eextra)_{s}$) 
and background ($F(\eextra)_{b}$)
simulations. The \eextra\ distribution
for signal events is modeled as the sum of an exponential and
two Gaussian distributions. The
double-Gaussian models signal events where the $X$ in $\btodlnux$ is a
\piz\ or photon with a characteristic energy around $0.15\gev$. 
The exponential models signal events
where such neutral particles are absent.
To model background, as determined from Monte Carlo, 
we use a third-order polynomial. The \PDF's are
combined into an extended maximum likelihood function,
\begin{equation}
\calL(s+b) \equiv \frac{e^{-\mu_s-\mu_b}}{n!} \prod_{i=1}^{n}\left[ \mu_s F(E_{i})_{s} + \mu_b F(E_{i})_{b} \right],
\end{equation}
where $E_{i}$ is the \eextra\ in the $i$th event,
$n$ is the total number of events in the data, and $\mu_s$ and $\mu_b$ are the signal
and background yields to be fitted in the data. Studies of the
choice of \PDF\ parameterization and of variations in shape
suggest that the chosen \PDF's yield a consistently
conservative limit for the upper bound of the branching
fraction. We fix the \PDF\ shape parameters and 
fit the data (Fig. \ref{fig:sl_eextra}). The fit yields 
$14.8 \pm 6.3$ signal events
and $115.2 \pm 11.8$ background events. This signal yield
has a statistical significance of $2.3\sigma$.

We set a limit on the branching fraction at the 90\% confidence level (C.L.) 
using the ``CLs method'' described in Refs.~\cite{bib:cls}\cite{bib:lepcls}. 
We define our statistical estimator, $Q$, to be the fitted signal yield 
and compare the value of $Q$ in data to its value in a large
number of experiments generated by sampling the likelihood function
over a range of signal hypotheses.
The uncertainty in the signal efficiency estimate is included by
assuming a Gaussian uncertainty in the signal hypothesis.
Using our fitted signal yield, efficiency, and the total
number of \B\ mesons in the data sample we determine that  
$\BRBtaunu < 6.7 \times 10^{-4} \, \textrm{(90\% C.L.)}$.


In our second method, we reconstruct the tag \B\ candidate, $\B_{{\rm had}}$, 
decaying into a set of purely hadronic final states,
$\Bp \to \bar{D}^{(*)0} X^+$. The $\bar{D}^{*0}$\
is reconstructed in the mode $\Dzb\piz$, and $X^+$ is a 
system of hadrons composed of
$n_1 \pipm + n_2 \Kpm + n_3 \piz + n_4 \KS$ where
$n_1=1,...5$; $n_2=0,1,2$; $n_3=0,1,2$; and $n_4=0,1$.
Rejection of background processes 
is based on two kinematic quantities: \DeltaE, the difference between
the $\B_{{\rm had}}$ and beam energies,
and the beam-energy-substituted mass \mes,
\begin{equation}
\mes \equiv  \sqrt{[(s/2+ \vec{p}\cdot{\vec{p}_\B})^2 / E^2 ] - |
\vec{p}_{\B}|^2} \: ,
\end{equation}
where $\sqrt{s}$ is the total energy of the \epem\ system in the
CM frame, and $(E, \vec{p})$ and $(E_\B, \vec{p}_{\B})$ 
are the four-momenta of the \epem\ system and the $\B_{{\rm had}}$,
respectively, both in the laboratory frame.

\begin{figure}[!t]
  \begin{center}
    \includegraphics[width=\linewidth]{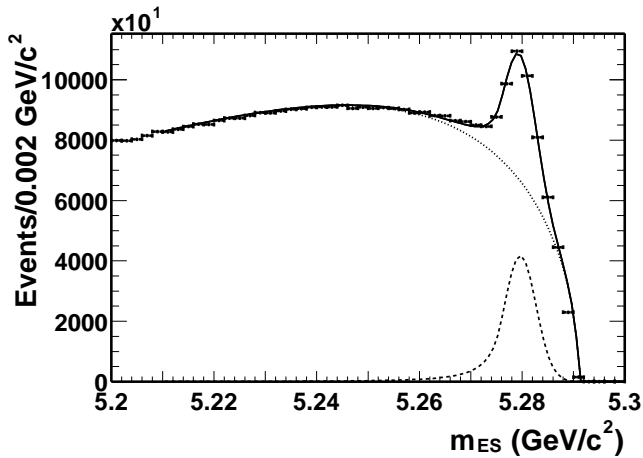}
    \caption{Distribution of \mes\ for the $\B_{{\rm had}}$ candidates
      in data. The events lie in the region $ -0.1 < \DeltaE <
0.08 \gev $.      
      The solid curve shows the result of the fit with the sum of a
      Crystal Ball function (dashed curve) and an ARGUS function (dotted curve). 
    }
    \label{fig:mesfit}
  \end{center}
\end{figure}

For each mode the \mes\ distribution
of the reconstructed candidates with $ -0.1 < \DeltaE <
0.08 \gev $ and $\mes > 5.21 \gevcc$ is fitted using the sum of a
``Crystal Ball function''~\cite{bib:cb} to model the signal component peaking 
at $m_{B}$ and an ``ARGUS function''~\cite{bib:argus} to model 
the continuum and combinatorial \B\ background.
Figure~\ref{fig:mesfit} shows the fit to the \mes\ distribution for the 
$\B_{{\rm had}}$ candidates in data.
We define the signal region as $ -0.09 < \DeltaE <
0.06 \gev $ and $\mes > 5.27 \gevcc$. We define a {\em sideband} region,
$ 5.21 < \mes < 5.26 \gevcc$, to provide a control sample for studying
continuum and combinatorial \B\ background.
The yield in the signal region, as determined
from the fit, is $N_{B_{{\rm had}}}= (167.8 \pm 1.2_{stat} \pm 3.0_{syst}) \times 10^3$. 
The error is dominated by systematic uncertainty in the functional
form of the peak at $m_{B}$.

We identify the $\tau$ lepton using the following decay channels: 
$\taumtoe$, $\mun\nut\numb$, $\pim\nut$, $\pim \piz \nut$,
and $\pim \pip \pim \nut$. 
We require the charged particles to be identified as leptons or pions, 
as appropriate. Mode-specific constraints are placed on 
the particles recoiling against the $B_{\rm had}$.
For the lepton and single-pion modes 
we reject events with $\piz$ or $\KS$ mesons in the recoil.
The event is required to have zero charge and, in the recoil,
at most one photon candidate not associated with a \piz. Events
with such a photon candidate are accepted only if 
$50<E_{\gamma}<100\mev$ ($50<E_{\gamma}<110 \mev$ for the $\taumtoe,
\mun\nut\numb$, and $\pi\nut$ modes) in the laboratory frame.
Further requirements are made on the total missing momentum of the event,
$p_{\mathrm{miss}}>1.2\gevc$ ($>1.4\gevc$ for $\taum \to \pim \piz \nut$),
the total momentum of the track(s) in the parent-\B\ rest frame
($p_{\pim}>1.2\gevc$ for \taumtopi, $p_{\pim \pip \pim}>1.6\gevc$ for
$\taum \to \pim \pip \pim \nut$), and the invariant mass of two or three
pions ($0.60<m_{\pi\pi}<0.95\gev/c^2$ and  $1.10<m_{\pi\pi\pi}<1.60\gev/c^2$ for
$\taum \to \pim \pip \pim \nut$, $0.50<m_{\pim\piz}<1.00\gev/c^2$ for
$\taum \to \pim \piz \nut$).

We use detailed Monte Carlo simulations to determine for each $\tau$
decay channel the selection 
efficiencies $\varepsilon_i$  weighted by the
corresponding branching fractions~\cite{bib:pdg2002}.
The systematic uncertainties in selection efficieny arise from tracking efficiency,
neutral reconstruction, particle identification, and \piz\ reconstruction.
The total \Btaunu\ selection efficiency (see Table~\ref{tab:semiexclsummary})
is $(10.5 \pm 0.2) \%$. Misreconstruction and
contamination amongst the $\tau$-decay channels are taken into account.

Continuum and combinatorial \B\ background is determined by
extrapolating the ARGUS function from the \mes\ sideband into the
\mes\ signal region.
The background that peaks in the \mes\ signal region 
is determined from Monte Carlo simulations
of \BpBm\ events. Events where a \Bz\ is incorrectly reconstructed
as a \Bp\ provide a negligible contribution.

We correct the expected background, $b_i$, to
take into account possible dependencies of the fitted
ARGUS shape on a given discriminating variable ($p_{\rm miss}$,
invariant masses, etc.). The correction factor is the ratio of the background 
expectations determined using two separate methods.
In the first method, we estimate the background 
by scaling the number of events in the \mes\ sideband using the
ARGUS signal-to-sideband ratio.
In the second method, we bin
each discriminating variable and reweight
the number of events, bin-by-bin, using
the ARGUS signal-to-sideband ratio for each bin.
The systematic error on $b_i$
is estimated as the deviation from unity
of the total correction factor for each $\tau$-decay mode.
The expected background and the total systematic uncertainty in
each $\tau$-decay channel is reported in Table~\ref{tab:semiexclsummary},
along with the number $n_i$ of selected candidates in data.

The systematic uncertainty in $N_{B_{{\rm had}}}$ ($1.8\%$) is estimated 
as the change in the yield in the signal region in Fig.~\ref{fig:mesfit}
when we use a double Gaussian as an alternative to the Crystal Ball function.
Other models for the signal or the background distribution
result in  negligible changes.

We observe a total of 15 \Btaunu\ candidates, which is consistent with
the expected background of $17.2 \pm 2.1_{{\rm stat}} \pm 1.3_{{\rm
    syst}}$ events. The distribution of these events is also
consistent with background.

We determine the \Btaunu\ branching fraction from the number of signal
candidates $s_i$ expected for each $\tau$ decay mode, where $s_i \equiv
N_{\B_{{\rm had}}} \BRBtaunu \varepsilon_i$.
The results for each decay channel are combined using the estimator, $Q$.
Here we define $Q$ to be ${\cal L}(s+b)/{\cal L}(b)$, where
\begin{equation}
  {\cal L}(s+b) \equiv
  \prod_{i=1}^{n_{{\rm ch}}}\frac{e^{-(s_i+b_i)}(s_i+b_i)^{n_i}}{n_i!},
        \;
  {\cal L}(b)   \equiv
  \prod_{i=1}^{n_{{\rm ch}}}\frac{e^{-b_i}b_i^{n_i}}{n_i!}
  \label{eq:lb}
\end{equation}
are the likelihood functions for signal-plus-background and background-only
hypotheses and  $n_{{\rm ch}}$ is the total number of reconstructed $\tau$-decay channels.

\begin{table}[t]
\caption{\label{tab:semiexclsummary}
Branching fraction ($\mathcal{B}$)\cite{bib:pdg2002}, 
efficiency ($\varepsilon_i$), expected background ($b_i$) with statistical and systematic errors,
and observed data candidates ($n_i$) for each reconstructed $\tau$
decay mode.
}
\begin{center}
\begin{tabular}{l c c c c} \hline \hline
selection         & $\mathcal{B} (\%)$ & $\varepsilon_i (\%)$ &  $b_i$                     & $n_i$ \\ \hline
$e\nu\nu$         & $17.84 \pm 0.06$   & $\:\:3.4 \pm 0.1$ &  $\:\:0.7 \pm 0.4 \pm 0.1$ & $\:2$\\ 
$\mu\nu\nu$       & $17.37 \pm 0.06$   & $\:\:1.9 \pm 0.1$ &  $\:\:0.9 \pm 0.5 \pm 0.1$ & $\:0$\\ 
$\pi\nu$          & $11.06 \pm 0.11$   & $\:\:2.6 \pm 0.1$ &  $\:\:1.3 \pm 0.6 \pm 0.2$ & $\:2$\\ 
$\pim\pip\pim\nu$ & $\:9.52\pm 0.10$   & $\:\:0.6 \pm 0.1$ &  $\:\:4.3 \pm 1.0 \pm 0.3$ & $\:4$\\ 
$\pim\piz\nu$     & $25.41 \pm 0.14$   & $\:\:2.0 \pm 0.1$ &  $10.0 \pm 1.6 \pm 1.3$    & $\:7$\\ 
\hline
all               & $81.20 \pm 0.22$   & $10.5 \pm 0.2$    & $17.2 \pm 2.1 \pm 1.3$     & $15$ \\
 \hline \hline
\end{tabular}
\end{center}
\end{table}
Since we have no evidence of signal we set an upper limit on
the branching fraction.
The statistical and systematic uncertainties in the expected
background are included in the estimator $Q$ by convolving 
the likelihood functions with a Gaussian distribution
having as standard deviation the combined statistical and
systematic errors  in the background estimate~\cite{bib:lista}. 
We determine that 
$\BRBtaunu < 4.2 \times 10^{-4}$~($90\%$ C.L.).


To combine the results from the
statistically independent hadronic and semileptonic samples,
we first calculate the likelihood ratio estimator, 
$Q \equiv \calL(s+b)/\calL(b)$, using the likelihood functions 
from each method. We create a combined estimator from the product
of the semileptonic ($Q_{{\rm sl}}$) and 
hadronic ($Q_{{\rm had}}$) likelihood ratio
estimators, $Q = Q_{{\rm sl}} \times Q_{{\rm had}}$.
The measured branching fraction, which is the value that maximizes 
the likelihood ratio estimator, is $(2.3^{+1.5}_{-1.3})\times 10^{-4}$.
The lower one-standard-deviation bound does not include zero because 
of the small excess of signal events observed in the 
semileptonic analysis.
Since this value is compatible with a zero branching fraction,
we set a combined upper limit,
\begin{equation}\label{eqn:thelimit}
\BRBtaunu < \thelimit \, \textrm{($90\%$ C.L.).} 
\end{equation}
The semileptonic
analysis does not contribute significantly to the combined limit
because of the observed small excess of signal events.

We use Eq.~\ref{eq:brsm}, Eq.~\ref{eqn:thelimit}, and
the measured value of $|V_{ub}|$ to set a limit on \fsubb. We
find $f_{B}<0.510\gev$ (90\% C.L.).

In conclusion, we have searched for \Btaunu\ in the recoil of
hadronic and semileptonic \B\ decays. We have set the most stringent
upper limit to date on this process.

We are grateful for the excellent luminosity and machine conditions
provided by our \pep2\ colleagues, 
and for the substantial dedicated effort from
the computing organizations that support \babar.
The collaborating institutions wish to thank 
SLAC for its support and kind hospitality. 
This work is supported by
DOE
and NSF (USA),
NSERC (Canada),
IHEP (China),
CEA and
CNRS-IN2P3
(France),
BMBF and DFG
(Germany),
INFN (Italy),
FOM (The Netherlands),
NFR (Norway),
MIST (Russia), and
PPARC (United Kingdom). 
Individuals have received support from CONACyT (Mexico), A.~P.~Sloan Foundation, 
Research Corporation,
and Alexander von Humboldt Foundation.


\begin{thebibliography}{99}

\bibitem{bib:cc}
Charge-conjugate modes are included implicitly throughout this paper.

\bibitem{bib:CKM}
N.~Cabibbo, \jprl{10}, 531 (1963);
M.~Kobayashi and T.~Maskawa, \progtp{49}, 652 (1973).

\bibitem{bib:pdg2002}
Particle Data Group, S. Eidelman {\em et al.}, \plb{592}, 1 (2004).

\bibitem{bib:higgs}
W.-S.~Hou, \jprd{48}, 2342 (1993).

\bibitem{bib:L3}
L3 Collaboration, M.~Acciarri {\em et al.}, \plb{396}, 327 (1997).

\bibitem{bib:babar}
\babar\ Collaboration, B.~Aubert {\em et al.}, \nima{479}, 1 (2002).

\bibitem{bib:geant}
{\geant}4 Collaboration, S.~Agostinelli {\em et al.}, \nima{506}, 250 (2003).

\bibitem{bib:foxwolfram}
  G. C. Fox and S. Wolfram, \jprl{41}, 1581 (1978).


\bibitem{bib:cls}
A. L. Read, ``Presentation of Search Results: The CL(S) Technique.'',
\jpg{28}, 2693 (2002).

\bibitem{bib:lepcls}
ALEPH Collaboration and DELPHI Collaboration and L3 Collaboration and 
OPAL Collaboration and LEP Working Group for Higgs boson searches,
R. Barate {et al.}, \plb{565}, 61 (2003).



\bibitem{bib:cb}
E. Bloom and C. Peck, \arnps{33}, 143 (1983);
Crystal Ball Collaboration, D. Antreasyan, Crystal Ball Note 321 (1983).
 
\bibitem{bib:argus}
ARGUS Collaboration, H.~Albrecht {\em et al.}, \plb{185}, 218 (1987).

\bibitem{bib:lista}
L.~Lista, \nima{517}, 360 (2004).

\end{thebibliography}
\end{document}